\documentclass{statsoc}
\usepackage[a4paper]{geometry}
\usepackage{graphicx}
\usepackage{amsmath}
\usepackage{multirow}
\usepackage{hyperref}

\usepackage{natbib}
\usepackage[absolute]{textpos}
\title[Polling bias and undecided voter allocations]{Polling bias and undecided voter allocations: US presidential elections,  2004--2016}

\author[Bon {\it et al.}]{Joshua J Bon}
\address{
School of Mathematics and Statistics,
University of Western Australia,
Perth, WA,
Australia.}
\email{joshuajbon@gmail.com}
\author{Timothy Ballard}
\address{
School of Psychology,
University of Queensland,
Brisbane, QLD,
Australia.}
\author[Bon {\it et al.}]{Bernard Baffour} 
\address{
School of Demography,
Australian National University,
Canberra, ACT,
Australia.}

\begin{document}
\begin{textblock}{18}(1,1)
\noindent Please cite as: \newline
Bon, J. J., Ballard, T. and Baffour, B. (2019), Polling bias and undecided voter allocations: US presidential elections, 2004--2016. Journal of the Royal Statistical Society, Series A (Statistics in Society), 182(2): 467-493. \url{http://dx.doi.org/10.1111/rssa.12414}
\newline\newline
This is the author accepted manuscript.
\end{textblock}

  \begin{abstract}
    Accounting for undecided and uncertain voters is a challenging issue for predicting election results from public opinion polls. Undecided voters typify the uncertainty of swing voters in polls but are often ignored or allocated to each candidate in a simple, deterministic manner. Historically this may have been adequate because the undecided were comparatively small enough to assume that they do not affect the relative proportions of the decided voters. However, in the presence of high numbers of undecided voters, these static rules may in fact bias election predictions from election poll authors and meta-poll analysts. In this paper, we examine the effect of undecided voters in the 2016 US presidential election to the previous three presidential elections. We show there were a relatively high number of undecided voters over the campaign and on election day, and that the allocation of undecided voters in this election was not consistent with two-party proportional (or even) allocations. We find evidence that static allocation regimes are inadequate for election prediction models and that probabilistic allocations may be superior. We also estimate the bias attributable to polling agencies, often referred to as ``house effects''.
  \end{abstract}


\section{Introduction}
Timely and accurate polls are crucial in describing current political sentiment and trends. Whilst no one poll will be sufficiently precise to enable reliable election predictions, combining the results of many pre-election polls has traditionally been viewed as a way to provide accurate forecasts.  However, bias at the level of the individual poll can produce systematic error in aggregate results, particularly if these biases are correlated. One important source of polling bias arises from undecided voters. For this reason, the eventual accuracy of pre-election polls is influenced by what is done to those respondents who are undecided. In the 2016 US presidential election, a large share of voters remained indecisive up until election day. When large in number, likely voters uncertain in their candidate preferences have the power to determine tight elections. Most polling firms deal with undecided voters using deterministic allocation methods, the most popular being proportional or equal allocation. Static allocation methods prevent the uncertainty attributable to undecided voters from propagating through a model, and may contribute to systematic bias in the polls when the undecided voters do not split as assumed. For this reason, the use and study of probabilistic allocation methods for undecided voters is an important, but as yet under-researched area.

\begin{figure}
    \centering	\makebox{	\includegraphics[width=0.8\textwidth]{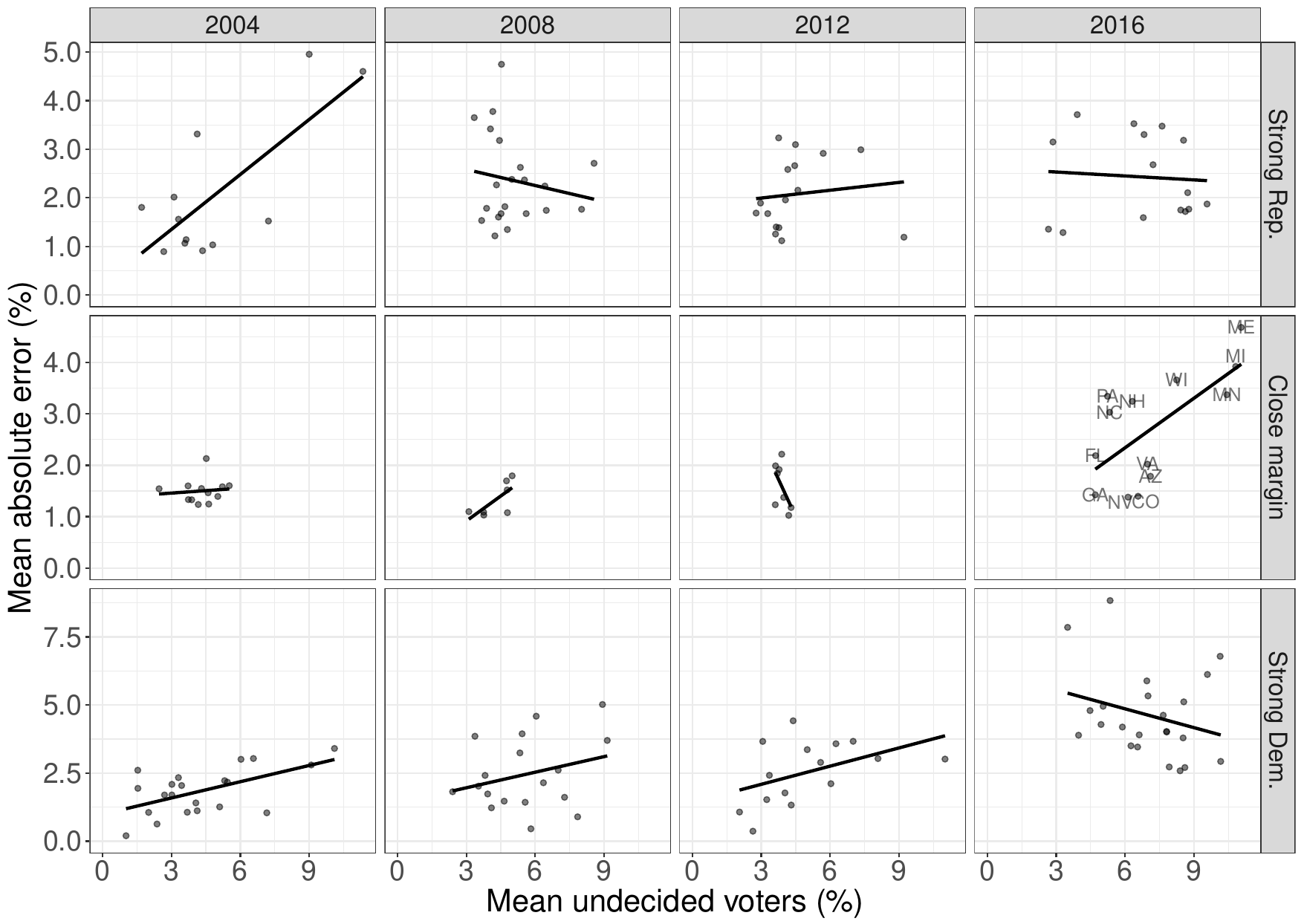}
	}
    \caption{\label{fig:und-err-grid}Mean absolute error of state polls versus mean level of undecided voters in each state, separated by year and margin of state election victory. Polling data are polls within 35 days of US presidential elections from 2004 to 2016. ``Close margin'' categorises state-level elections with absolute margin $\leq6$\%, ``Strong Rep.'' are races where the Republican candidate had margin$>6$\%, and ``Strong Dem.'' is the remainder.}
\end{figure}

 From a polling perspective, the 2016 US presidential election is of interest because of the public perception (and media narrative) of polling failure and the impact of high levels of undecided voters in the lead up to election day \citep{aapor2016}.  Figure~\ref{fig:und-err-grid} shows the relationship between each state's absolute polling error and undecided voters grouped by year (2004 to 2016) and election result margin (a strong Republican, close, or strong Democrat victory). The centre row is of most interest, as relatively high errors have the most impact in elections where the race is close. Within this subset, 2016 shows the strongest association between mean absolute error and undecided voters, indicating that the role of the undecided may have been unprecedented in the 2016 presidential election.
 
	In this paper, we investigate the impact of undecided voters in the 2016 presidential election and presidential elections in recent years. We begin by providing background information on surveys, election polling, and undecided voters. Next, we introduce the data on national undecided voter levels to motivate our interest in the 2016 election, followed by state-level polling data for our analysis. We present a novel extension of a recently proposed model \citep{Shirani2018} that allows us to quantify the bias attributable to undecided voters. We are also able to include bias attributable to certain polling agencies, often referred to as ``house effects''. We show that bias in the 2016 US presidential election was critically higher than in the previous presidential elections, and a sizeable proportion of this increase can be accounted for by the high levels of undecided voters. Finally, we discuss our conclusions and recommendations from this work. 
	
\section{Surveys, election polling and undecided voters}
\subsection{Surveys and election polling}
The accuracy (or lack thereof) of public opinion surveys and election polls has received substantial attention in recent years. All public opinion surveys, hence polls, are predicated on the assumption that citizens possess well developed attitudes on major political issues, and that surveys are passive measures of these attitudes \citep{Converse1986,Zaller1992}. In practice however, surveys may fail to adequately capture the sociological complexity of voting decisions and behaviour \citep{Crossley1937,Gelman1993,Jacobs2005,Hillygus2011}.  Increasingly, modern public opinion surveys also have to cope with declining response rates \citep{Keeter2000,Tourangeau2013,PRC2012} combined with difficulties in achieving complete coverage of the population \citep{Jacobs2005,Traugott2005,Leigh2006,Erikson2008}. Robust evidence has demonstrated that the role of statistical uncertainty in the opinion polls has not been adequately understood \citep{Martin2005,Sturgis2016,Hillygus2011,Graefe2014} and a failure to fully reflect this uncertainty leads to an over-statement of confidence level in predictions from survey outcomes \citep{Erikson2008,Rothschild2015,Lock2010}.

There are a number of factors that influence the uncertainty in election polls. First, although polls play an important role in the democratic process it is becoming increasingly difficult to measure voting intention \citep{Curtice2008,Keeter2016}. Relatedly, there are differences between voting intentions and voting behaviour \citep{Wlezien2013,Hopkins2009,Veiga2004,Jennings2016}. It is also generally agreed that survey respondents will not be fully representative of the entire voting public. Finally, disparities in the population result in differences in voter behaviour by geography, ethnicity, social class, gender and age, which effectively exacerbates the level of uncertainty when it comes to generalising from the sample survey to the broader population.

Understanding the mechanisms that affect polling error can provide valuable insights for better calibration and efficiency of individual polls and models derived from them. Yet undecided voters are still a relatively under-studied source of this error. 

\subsection{Undecided voters and election polling}
Pre-election polls are typically conducted based on random sampling of likely voters who are asked for their preference among presidential candidates. Most polls record the percentage of voters who respond indecisively but do not necessarily publish this level in their survey results \citep{Sturgis2016}. When the number of undecided voters is not reported, the polling agency may simply pre-allocate based on some judgement or decision-rule (which they may or may not report). Conversely, when the level of undecided voters is reported it can still be a challenge to incorporate these into predictive models \citep{Hillygus2011,Hoek1997}. There is no consistent or transparent method for handling undecided voters.

To begin, we define undecided voters as individuals that are likely to vote but who have not formed a voting intention when surveyed prior to election day. The term ``late-deciding'' also describes these voters. Whilst similar to \citet{Kosmidis2010}, our definition is restricted to likely voters as most election polls make adjustments to report results for this group only  \citep{Sturgis2016}. Moreover, by necessity our definition of undecided voters must also include likely voters who have chosen not to disclose their voting preferences by stating they are undecided during a survey.

Many rule-based methods have been proposed to handle undecided voters in elections \citep{Crespi1988,Daves1995,Fenwick1982}, however some findings have indicated these assignment methods do not improve forecast accuracy \citep{Hoek1997}. Simple rules for allocating undecided respondents may be adequate if the undecided voters are small in number but will likely fail when these numbers are relatively high. Additionally, any deterministic rule will not allow for variability of allocations to be modelled in predictive outcomes, which is problematic for statistical calibration. Limited research into the impact of undecided voter allocation on election poll modelling still leaves many questions about the role of indecisive voters in election polling.

The effect of undecided voters on the assessment of predictive accuracy of election polls has been considered previously \citep[see for example,][]{Mitofsky1998,Hoek1997,Visser2000,Martin2005}. Overall, the research has focussed on the treatment of undecided voters so that consistent accuracy measures can be defined, rather than how allocative assumptions impact polling bias. \citet{Visser2000} state that there is little published ``collective wisdom'' on undecided voters and better guidelines are needed, especially since excluding undecided voters was the least effective strategy in their analysis.

Investigation into undecided voter behaviour has occurred mostly in the context of election campaign assessment. For example, in US and Canada, voters who decide last minute may be more open to persuasion \citep{Chaffee1996,Fournier2004}, and in the 2005 British elections, \citet{Kosmidis2010} concluded that perceived economic competence was a larger driver for the behaviour of undecided voters. However, for election outcome modellers, little can be said on the correct treatment of undecided voters for election predictions.  Most notably, imputing candidate preferences for undecided voters has been found to be somewhat beneficial in \citet{Fenwick1982} whilst more recently \citet{Nandram2008} proposed a Bayesian allocation. However, the true benefits for election forecasting is not yet clear.

Formal reports into the polling of the most recent elections in the US and UK have found some evidence of bias attributable to undecided voters. In the 2015 UK general election, \citet{Sturgis2016} report a modest, but marginal, effect from late-deciding voters, at most 1\%. They assign the primary cause of the polling failure to unrepresentative samples, which statistical procedures (designed to account for this) were not able to mitigate. The report into polls for the US presidential election \citep{aapor2016} suggest that polls were accurate at the time they were conducted, but in some key states projection error was high due to late-deciding voters. Overall, they attribute a number of factors to polling bias in the 2016 election. In addition to late or undecided voters, they particularly emphasise over-representation of college graduates (without appropriate adjustment) and late-revealing Trump supporters -- which can also manifest as larger levels of undecided voters in polls. 

Based on the accuracy of the 2016 presidential election, a more sophisticated evaluation of undecided voters is necessary for predicting future elections if undecided voter levels are relatively high. This study estimates the sizeable bias attributable to undecided voters in 2016, shows evidence of undecided bias in previous presidential elections, and demonstrates that allocating undecided voters in proportion (or evenly) to the leading candidates is a poor assumption. We expect the implications of our findings will contribute to uncovering causal mechanisms that (probabilistically) determine undecided voters allocations. Unfortunately, with the available data we are not yet able to investigate these. We elaborated on the consequences for election prediction and future research in the discussion. 

\subsection{Meta-analysis of polls}
Any one poll will be a snapshot of the sample collected, fraught with difficulties pertaining to sampling design, non-representativeness, and differences in methodological assumptions. For this reason, a single poll should be interpreted cautiously, even more so in situations when there is little previous experience to draw upon, for example in a referendum or, arguably, the 2016 US presidential election. Nonetheless, polling results from various sources can be compiled, compared, analysed and then interpreted using meta-analysis techniques to combine (or pool) together different polls.

Meta-polls and poll modelling can compensate for the bias and inaccuracy of individual polls, but establishing well-calibrated models require understanding the inherent problems in polls, and defensible model assumptions. In election polling, deterministic (rule-based) allocation of undecided voters is widespread \citep{Crespi1988,Visser2000,Martin2005}. These methods are appealingly simple and create a consistent set of data when polling organisations do not publish the number of undecided or third party voters. Undecided voters can be allocated (explicitly or implicitly) in a number of ways \citep{Martin2005, Mitofsky1998}. The most prevalent are:
\begin{itemize}
	\item Splitting the undecided voters proportionately between the two leading candidates. This is equivalent (in mean) to discarding the undecided voters and normalising the two leading candidate's voter proportions, and
	\item Allocating half of the undecided voters to each of the leading candidates. This is equivalent (in mean) to only reporting the margin between the two leading candidates.
\end{itemize}
Identifying the allocation procedures that polling firms use (if they do not report undecided voters) is difficult because they are averse to providing commercially sensitive information. Some meta-pollsters have published how they handle undecided voters in their models for at least the 2016 election. FiveThirtyEight split undecided evenly between the major-party candidates \citep{Silver2016}, as did the Princeton Election Consortium implicitly when they used the margin between the two leading candidates \citep{PEC2016}. In contrast, the Huffington Post used a different strategy, assuming that ``one-third of undecided voters won't vote; one-third will gravitate nationally toward either candidate; and the remaining one-third will add to this state's margin of error'' \citep{HP2016a}. Interestingly, the Huffington Post model is a mixture between proportional allocation, even allocation, and (imprecisely) incorporating some uncertainty from undecided voters into poll modelling.

Meta-analysis of polls is crucial to obtaining reliable predictions for elections, however, it is very difficult for these models to account for systematic bias in polls. Namely, if every poll is subject to a particular source of bias, how do we isolate and quantify its influence without external information? Investigating the role of undecided voters in polling bias will help to understand one aspect that contributed to larger polling bias in the 2016 presidential election, which may occur again.

\subsection{Data}
In Section~\ref{und-national-analysis}, we examine the extent to which 2016 was an abnormal presidential election by considering the number of undecided voters relative to previous years. To investigate we summarise national polling data in US presidential elections from 2004 onwards, totalling 616 national polls. Publicly available polls that reported a sample size were used.  Polls from 2012 and 2016 were obtained from the Huffington Post's Pollster API \citep{HP2016b,Arnold2016}, data for 2008 were retrieved from an archived version of ``Pollster.com'' \citep{HP2009}, and data from 2004 were reconstructed with polls available from \citep{RCP2004}. Specifically, we reconstructed polls from RealClearPolitics that reported a third party candidate but did not sum to 100\% of the sample, and assumed that the undecided category was equal in size to the remaining proportion. In order to model the undecided voters, the majority of polls in each election year needed to report the undecided category (explicitly or implicitly). This limited the number of elections that could be analysed to 2004, 2008, 2012, and 2016. 

State level polling data from 2004 to 2016 are used to model polling bias and variance. The polls for 2012 and 2016 were obtained from \citet{HP2016b} whilst the polls for 2004 and 2008 were retrieved from US Election Atlas \citep{Leip2008}. Polls were included if they occurred up to 35 days prior to their respective election. Due to the complexity of our model, state-level elections were only included if they had at least 5 polls in the dataset. Whilst other poll repositories exist for 2004 and 2008 state-level election polling data, none consistently reported undecided voter counts. In total 1,905 polls, from 129 state-level election races were analysed. State-level polling data for US presidential elections where a majority of the polls included an undecided voter category were not found for years prior to 2004 by the authors.

\section{Comparison of undecided voters in US presidential elections from national surveys}\label{und-national-analysis}
\begin{figure}
    \centering	\makebox{	\includegraphics[width=0.8\textwidth]{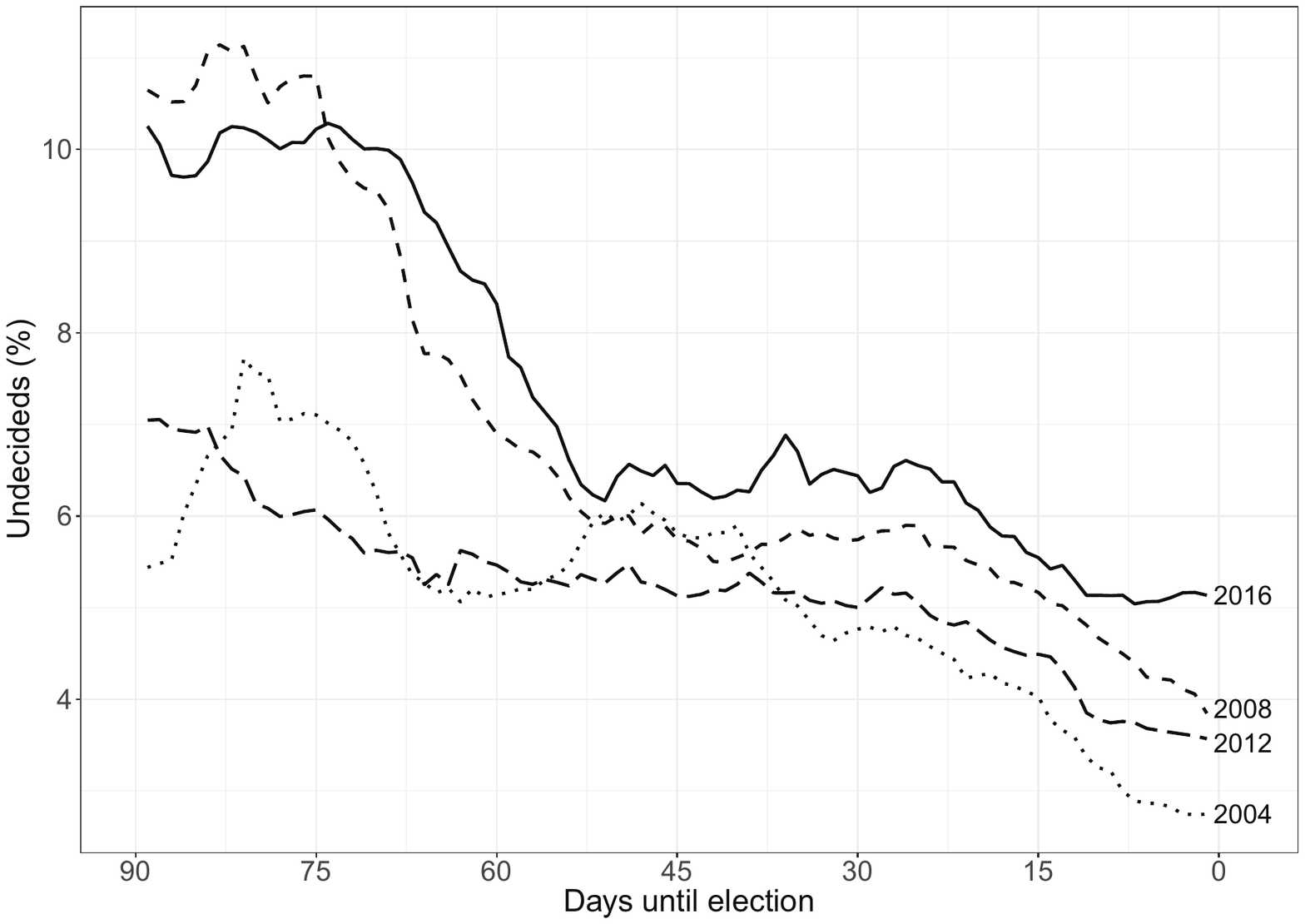}
	}
    \caption{\label{fig:und-roll-ts}Mean level of undecided voters, as captured by national polls, over the 90 days prior to US presidential elections from 2004 to 2016. The number of undecided voters on each day x is the weighted average from national polls that occur within a two-week window centred at x.}
\end{figure}

Undecided voters were much higher during the 2016 US presidential election relative to previous years. Figure~\ref{fig:und-roll-ts} shows the moving average number of undecided voters over the course of presidential elections from 2004 to 2016. It can be seen that the year 2016 had a larger number of undecided voters on average and that this trend was persistent over the course of the campaign.  Whilst 2016's pattern of undecided voters over time was similar to that of 2012, the major difference was that in the final week of polling the undecided voters did not continue to fall. It also appears that the 2004 and 2012 elections followed a similar pattern, however the extra variability in the 2004 election may be explained by lower numbers of polls and the reconstruction that took place. In the week prior to each election the weighted average of undecided voters was 5.1\%, 3.5\%, 3.9\%, and 2.7\% for 2016 to 2004 respectively.

\begin{figure}
    \centering	\makebox{	\includegraphics[width=0.8\textwidth]{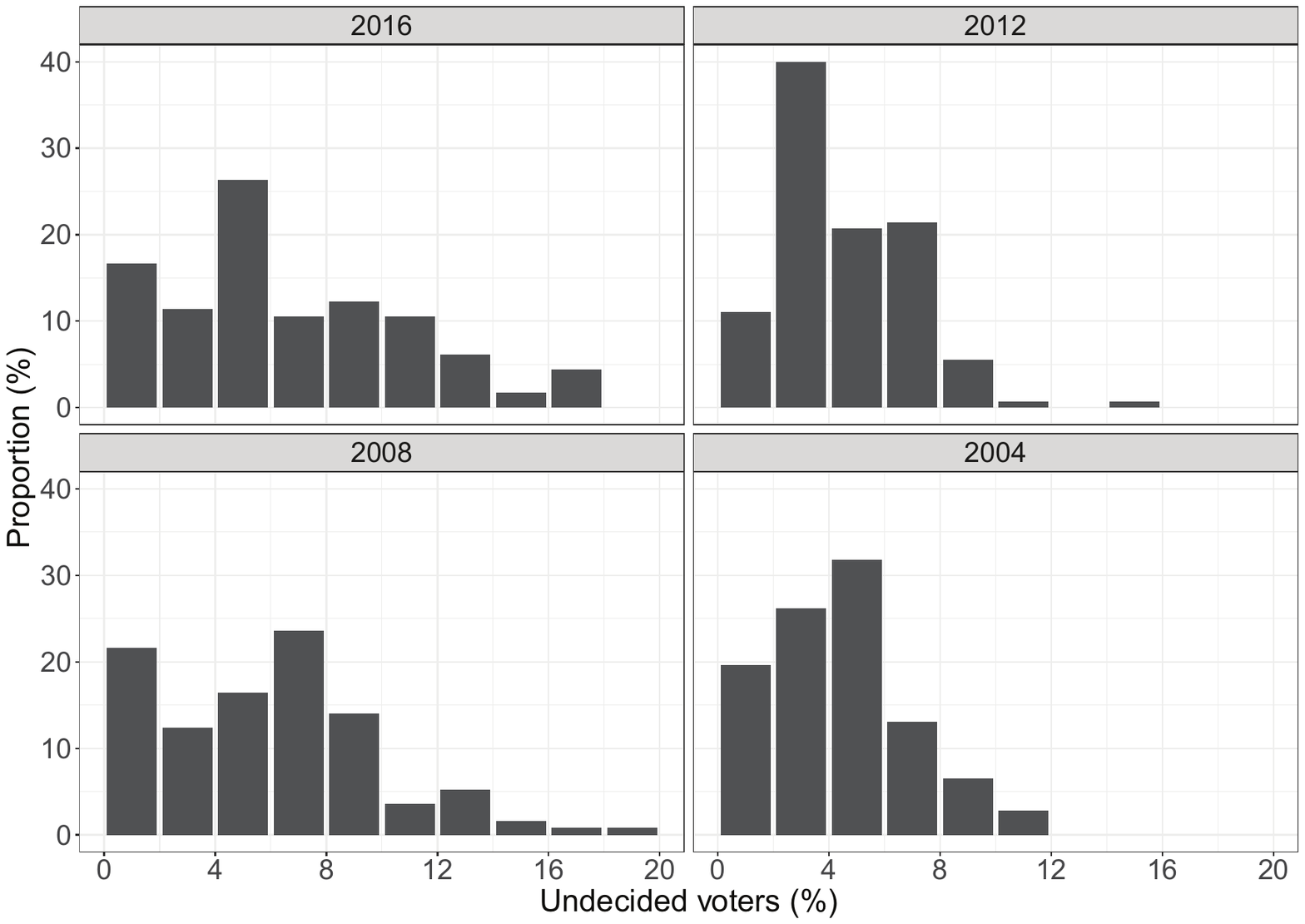}
	}
    \caption{\label{fig:und-dist}Histogram of undecided voters, as captured by national polls, over the course of 3 months prior to the US presidential elections 2004-2016. Each bar is relative to the number of polls from that year.}
\end{figure}

The distributions of undecided voters in the months leading up to presidential elections also appear to vary over time, as seen in Figure~\ref{fig:und-dist}. The undecided voters in 2016 and 2008 have relatively fatter tailed distributions, whilst the 2012 and 2004 elections appear to be centred between 3-4\%. Undecided voter levels higher than 10\% occurred more frequently in the 2016 election than 2012 and 2004, and somewhat more frequently than in 2008. The similarities of 2016 and 2008 as well as 2012 and 2004 may be partially attributable to the absence or presence of an incumbent candidate, but this is difficult to infer from only 4 elections. 

The levels of undecided voters in 2016 had an unusually high mean, differed more across states, and did not follow the final week decrease of previous elections. This finding, coupled with evidence from  Figure~\ref{fig:und-err-grid}, motivates an investigation into the effect of undecided voters on polling errors in the 2016 US presidential election.

\section{Methods}
\subsection{Assessing poll bias with the total survey error framework}
We use a total survey error framework \citep[see][for overview]{Biemer2010} to analyse state polls from the US elections. The paradigm has a long history, discussed as early as \citet{deming1944}. We refer readers to \citet{Groves2010} for a thorough introduction to the literature. The total survey error paradigm attempts to account for, and assess, many sources of error with respect to user requirements. In this analysis we take the user requirement to be predictive accuracy and calibration, and our users to be either polling organisations or poll aggregators. As such, our aim is to estimate bias and variance and describe the predictive characteristics of polls to these users. In particular, we quantify the bias and variance attributable to undecided voters, and demonstrate the inadequacy of proportional (or even) allocation rules.

As we are interested in ascertaining the impact of state-level influences, we consider each state-level election from each presidential election year to be a distinct election. With regard to modelling, henceforth an election refers to a state-level presidential election from an US election year, specifically, 2004, 2008, 2012 or 2016.

Under the total survey error framework, a survey error is defined as the deviation of a survey response from its true underlying value. This error can occur either through bias or variance. The bias term captures the systematic errors which are shared by all election polls, such as shared operational practices, infrastructure and sampling frames for example. The variance term captures sampling variation, and can account for variation due to different survey methodologies, software, statistical models, or survey weighting adjustments. Therefore, the poll error which is computed through comparing the election outcomes to the predictions from multiple election polls, can be decomposed into election level bias and variance terms. We adopt this framework to ensure that the analysis of undecided voters and their role in the bias is not conflated by non-sampling errors.

On the one hand, sampling errors arise from taking a sample rather than the whole population and are usually accounted for using standard survey sampling approaches, including post-stratification \citep{Holt1979}, calibration \citep{Deville1992}, imputation \citep{Gelman2002}.  On the other hand, non-sampling error is a catch-all term that refers to all other sources of error that are not a function of the sample. In theory, although a specific poll estimate may differ from the true election outcome, under favourable repeated sampling conditions polls should produce reliable estimates \citep{Assael1982}. However, in practice, it is well known that differences between poll results and election outcomes are only partially attributable to sampling error \citep{Ansolabehere1993}. Most statistical procedures to compensate for non-sampling errors assume near universal (or high) response, but this is far from the norm: the majority of election surveys have less than 10\% response rates \citep{PRC2012}. Further, in polling there is a general difficulty in measuring voting intention and voting behaviour because polls measure the beliefs and opinions of respondents at the time of the survey, and cannot fully capture what respondents will do on election day \citep{Bernstein2001,Silver1986,Rogers2014,Jowell1993}. Increasingly, efforts to mitigate against this can exacerbate the inaccuracy \citep{Ansolabehere2012,Voss1995,Gelman2016,Bailey2016}. This is especially true when it comes to dealing with those who are undecided, either because they are truly undecided, or are hiding extreme voting preferences \citep{Gerber2013}. The treatment of the group of polling responses that report to be undecided is therefore an important, yet relatively unstudied area of research.

\subsection{A Bayesian approach to total survey error incorporating undecided voters}\label{methods:model} 
We use a meta-analytic approach to compare the estimates from the individual polls to the eventual election outcomes. This combines information from the various polls to produce a pooled estimate for the differences between the state-election poll means and the election outcome -- which is our ground truth. The modelling framework is based on the model proposed by \citet{Shirani2018}. Their model quantifies the total error by estimating the vote share under a two-party preference, but does so by assuming proportional allocation of undecided voters. We extend their model to explicitly include undecided voter proportions. A second addition to our model is a term for house effects (or pollster-specific bias) of polling agencies. These two sources of bias do not account for all errors polls are subject to, but do provide a quantitative method for assessing how much variability in the voting outcome is due to these sources. For other sources, the model uses state-year net bias and variance terms to capture the remaining error in the polls.

Since there are relatively few numbers of polls in some elections, a simple measure, such as root mean squared error, may yield imprecise estimates of the election-level bias. We address this by fitting a Bayesian hierarchical latent variable model \citep{Gelman2007}. This method pools data to determine estimates of bias and variance in states with small numbers of polls, allows bias to vary over time, and better captures the variance in excess of that expected from a simple random sample.  The model ``borrows strength'' across states and time to estimate smoothed within state trends of both polling bias and undecided voters in each election. 

With small adjustments the notation in \citet{Shirani2018}, each poll is associated with an election denoted by the index $r[i]$. Let $y_{i}$ be the two-party support for the Republican candidate of poll $i$, $n_{i}$ be the sample size, and $t_{i}$ be the time at which the poll was conducted. Two-party support indicates that a proportional allocation of the undecideds voters (or scaling) has occurred, namely
\begin{equation} \label{prop-poll-alloc}
	y_{i} = \frac{R_{i}}{R_{i} + D_{i}}
\end{equation}
where the Republican and Democratic support as measured in poll $i$, is $R_{i}$ and $D_{i}$ respectively.
The time $t_{i}$ is the duration between the last day the poll was conducted and the relevant election date, and is scaled to be between 0 and 1. The Republican candidate's final two-party vote outcome is denoted by $v_{r}$. Each poll is assumed to be distributed by
\begin{align} \label{poll-model-1}
	\begin{split}
	y_{i} &\sim \mathcal{N}(p_{i}, \sigma^{2}_{i}) \\
	\text{logit}(p_{i}) &= \text{logit}(v_{r[i]}) + \alpha_{1r[i]} + t_{i}\beta_{1r[i]} \\
	\sigma^{2}_{i} & = \frac{p_{i}(1-p_{i})}{n_{i}} + \tau_{1r[i]}^{2}
	 \end{split}
\end{align}

where $\mathcal{N}(\mu,\sigma^2)$ denotes the normal distribution parametrised by mean and variance, and the first subscript (e.g. the $1$ in $\alpha_{1r}$) indicates that the parameter is part of the polling model. The vote, $v_{r}$, captures the true mean of the polls allowing election level bias to be estimated by $\alpha_{1r} + t_{i}\beta_{1r}$ on the logit scale. Estimation on this scale ensures the estimated poll value, $p_{i}$, is bound between 0 and 1. Bias on election day is $\alpha_{1r}$ and the time-varying bias coefficient is $\beta_{1r}$. As for the variance, $\tau^{2}_{1r}$ accounts for the excess variance above what is expected in a simple random sample. We assume a positive additive structure in the variance, but in theory it is possible to reduce variance below that of a simple random sample using stratification. In practice however, other errors due to nonresponse, measurement, and specification are likely ensure that the variance is always above that of a simple random sample. Additionally, \citet{Shirani2018} noted that a multiplicative variance structure gave qualitatively similar results to the additive variance assumption in their study.

The model is able to detect the bias in election polling at the state-year level by centring the model about the actual election outcome, whilst estimating the excess variance by anchoring the model variance at the level expected from a simple random sample. Elections with few polls are estimated by pooling the data across elections using hierarchal priors (see Table~\ref{tab:priors} in Appendix~\ref{priors}).

Using the estimated two-party support from polls, as in \eqref{prop-poll-alloc}, the model implicitly assumes that undecided voters are distributed proportionately to the major-party candidates. We relax this assumption by explicitly distributing the undecided voters in proportion but with flexibility. To illustrate, let the undecided voters from poll $i$ be $U_{i}$, with $R_{i}$ and $D_{i}$ as in \eqref{prop-poll-alloc}. Scaling the polls to exclude third party candidates, we assume that 
\begin{equation} \label{scale-3p}
	y_{i}^{\prime} = \frac{R_{i} + \lambda U_{i}}{R_{i} + D_{i} + U_{i}}
\end{equation}
where $0\leq \lambda \leq 1$ allocates the undecided voters to the Republican candidate. Rather than using proportional allocation, $\lambda = \frac{R_{i}}{R_{i} + D_{i}}$, as is the case in model \eqref{poll-model-1}, we use 
\begin{equation} \label{uncertain-alloc}
	\lambda = \frac{R_{i}}{R_{i} + D_{i}} + \theta_{i}
\end{equation}
where $\theta_{i}$ is an unknown bias (away from proportional allocation) which occurs at some level (i.e. poll, election or election year). Simplifying equation \eqref{scale-3p} with \eqref{uncertain-alloc} leads to the identity
\begin{equation}\label{prop-alloc-identity}
	y_{i}^{\prime} = y_{i} + u_{i}\theta_{i}
\end{equation}
where $y_{i}$, represents the Republican two-party vote share as in \eqref{prop-poll-alloc} and $u_{i} = \frac{U_{i}}{R_{i} + D_{i} + U_{i}}$ represents the scaled proportion of undecided voters. This observation motivates changing model \eqref{poll-model-1} to include undecided voters as an explanatory variable. 

The term $\theta_{i}$ measures the bias away from a proportional two-party split. However, using poll-level undecided voters as an explanatory variable is problematic for two reasons; it is subject to measurement error (it is a survey estimate), and the level of undecided voters varies over time (see Figure~\ref{fig:und-roll-ts}). The latter issue may confound $\theta_{i}$ with estimates of the time-varying component of bias already in the model, $\beta_{1r}$. 

To address these concerns we propose a model for the undecideds so that election day undecided voter levels can be included in model \eqref{poll-model-1}, rather than the undecided numbers from each poll. The model of the undecided voters is
\begin{equation} \label{und-model}
	u_{i} \sim \mathcal{N}\left(
	\alpha_{2r[i]} + t_{i}\beta_{2r[i]}, 
	\tau^{2}_{2r[i]}
	\right)
\end{equation}
where $\alpha_{2r}$ estimates the election day level of undecided voters, the observed change in undecideds over time is accounted for by $\beta_{2r}$, and $\tau^{2}_{2r}$ measures the variability within each state-level election race $r$. From the undecided model, $\alpha_{2r}$ becomes an explanatory variable measuring the election day level of undecided voters in each race. 

In addition to bias from undecided voters, we also wish to account for house effects, the bias attributable to particular polling agencies and groups. As such, the categorical variable $\kappa_{h}$ is added to the model. The extended model can be written as
\begin{equation} \label{poll-model-2}
	\begin{split}
		y_{i} &\sim \mathcal{N}(p_{i}, \sigma^{2}_{i}) \\
	\text{logit}(p_{i}) &= \text{logit}(v_{r[i]}) + \alpha_{1r[i]} + t_{i}\beta_{1r[i]} - \alpha_{2r[i]}\gamma_{g[i]} + \kappa_{h[i]} \\
	\sigma^{2}_{i} & = \frac{p_{i}(1-p_{i})}{n_{i}} + \tau_{1r[i]}^{2}
	 \end{split}
\end{equation}
where $\gamma_{g}$ has replaced $\theta_{i}$ in \eqref{uncertain-alloc} and \eqref{prop-alloc-identity}, and is the effect of mean undecided voters on bias varying by the groups of state-years given in Figure~\ref{fig:und-err-grid}. The 12 groups, index by $g$, are formed by the cartesian product of election year (2004, 2008, 2012, and 2016) and the election result margin groups. The latter categorises how close election results were; strong Republican (margin $>6$\% in favour of Republican), close margin (absolute margin $\leq6$\%), and strong Democrat (margin $>6$\% in favour of Democrat). These groups are in line with Figure~\ref{fig:und-err-grid} and were chosen because results with a margin greater than 6\% are extremely unlikely to be affected by undecided voter allocation -- in the sense that (for the majority of states) the winner collects all of the state's Electoral College votes. Hence, the close margin group will have the most impact on the eventual outcome of the election. The groups are also a crude measure of partisan strength and influence in a state. 

The $\gamma_{g}$ coefficient is not estimated at the poll level because we now estimate (and allocate) at the state-year level of undecided voters on election day ($\alpha_{2r}$). Moreover, using the 12 groups, rather than a coefficient for each state-year, occurs because of identifiability issues with $\alpha_{1r}$. In addition to accounting for measurement error, using \eqref{und-model} also allows polls that do not report undecided voters (i.e. they have missing data) to be included in the model since only the state-level mean enters model \eqref{poll-model-2}. 

The house effects from different polling agencies or groups are modelled by $\kappa_{h}$. House effects are errors specific to a polling firm, for example they may arise from flawed survey or modelling methods or partisan prejudice. We include bias terms for 39 different polling agencies, indexed by $h$. Not all polls have an associated house effect, only those where at least 8 polls were in the dataset. 

To ensure that the final inference is not substantially affected by the choice of prior, we specify weakly informative priors following the previous work \citep{Shirani2018}. The hierarchal specification of the priors pull the bias and variance estimates of the poll towards the state's average in a given election year. However, the effect is related to the number of polls in the particular election year, and the overall distribution across all polls. For states with few polls (in a given year), the estimates can be inferred from information from other polls (across state and time). For the $\gamma_{g}$ and $\kappa_{h}$ we specify priors with shrinkage towards zero, this ensures the probability of overestimating these effects is low. The list of priors and further explanation can be found in Table~\ref{tab:priors} in Appendix~\ref{priors}.

We maintain a focus on the assumption of proportional allocation of undecided voters as it allows us to use polls that do not report an undecided voter number to still enter model~\eqref{poll-model-2}. However, we rerun the model with an even split of the undecided voters to each party for a more robust analysis. The results are very similar to model with proportional allocation (see Figure~\ref{fig:state-eld-bias} for example) and are discussed further in Appendix~\ref{split-50}.

Bayesian posterior sampling was conducted using an adaptive Hamiltonian Monte Carlo sampler \citep{betancourt2018} implemented in \texttt{Stan} \citep{Carpenter2017}. The analysis was facilitated by the statistical coding environment and language \texttt{R} \citep{rteam2017}, the \texttt{R} interface to \texttt{stan}, \texttt{rstan} \citep{rstan2018}, and diagnostics were provided by \texttt{shinystan} \citep{shinystan2016}. Since the undecided voter proportions are between 0\% and 10\% approximately, we multiply $\alpha_{2r}$ by 10 so that it is on the same scale as $t_{i}$. This improves computational performance when estimating the model. Testing showed negligible impact on the estimates.

\section{Results}\label{sec:results}
First, we compare our results from election years 2004 to 2016 with results obtained by \citet{Shirani2018} for 2000 to 2012 (the final column of Table~\ref{tab:model-1-avgs}) using the model specified in \eqref{poll-model-1} and the aforementioned paper.  In Table~\ref{tab:model-1-avgs}, the average absolute bias (bias from $\alpha_{1r} + t_{i}\beta_{1r}$) and election day bias ($\alpha_{1r}$) are both considerably higher in 2016 than in the previous three election years (see Appendix~\ref{average-bias} for details on calculating these quantities). The measures were at least 1.1 percentage points above previous years, having more than twice as much bias in the case of 2004 and 2008.  The increased bias in 2016 can explain the increase to the overall 2004--2016 average bias compared to that of 2000--2012. The yearly averages shown in \citet{Shirani2018} are also consistent with the results in Table~\ref{tab:model-1-avgs} (2004, 2008, and 2012). 

\begin{table}
	\caption{\label{tab:model-1-avgs}Average election-level absolute bias and average election-level standard deviation across state-elections in given year(s) from model \eqref{poll-model-1}. Values shown are posterior mean (s.d.) in percentage points.}
  	\centering
	\fbox{%
	\begin{tabular}{l|*{4}{c}|*{2}{c}}
		 & & & & & \multicolumn{2}{c}{\em Overall} \\ 
		 & 2004 & 2008 & 2012 & 2016 & 2004--2016 & 2000--2012 \\ \hline
		\multirow{2}{*}{Average absolute bias} & 0.9\% & 1.2\% & 1.4\% & 2.6\% & 1.7\% & 1.2\% \\
							  				& (0.11) & (0.10) & (0.11) & (0.10) & (0.06) & (0.07) \\[0.15cm] 
		\multirow{2}{*}{Average absolute election day bias} 	& 0.9\% & 1.2\% & 1.3\% & 2.4\% & 1.6\% & 1.2\% \\
							  				& (0.13) & (0.12) & (0.14) & (0.13) & (0.07) & (0.08) \\[0.15cm]  
		\multirow{2}{*}{Average standard deviation} 			& 2.2\% & 2.2\% & 2.1\% & 2.4\% & 2.3\% & 2.2\%  \\
							  				& (0.05) & (0.04) & (0.04) & (0.05) & (0.03) & (0.04) \\
	\end{tabular}}
\end{table}

Whilst the bias in election polls seems to have played a large role in the abnormality of the 2016 election year's polls, the average standard deviation appears to be consistent across time. The average standard deviation in 2016 was only 0.2\% above the next highest year. This is not a qualitative difference given the range of values is only 2.1\% to 2.4\% from 2004 onwards. The consistency in average standard deviation over time lends strength to the conclusion that individual polls are subject to approximately twice as much standard deviation than what is reported (i.e. a simple random sample calculation) \citep{Shirani2018,rothschild2016}. 

Second, Table~\ref{tab:model-2-avgs} lists the average results from model \eqref{poll-model-2} where several bias definitions are considered. Average absolute bias describes the average of election-level bias from all sources, whilst average absolute election day bias consists of all sources but the time-varying component (i.e. $t_{i} = 0$). Average absolute undecided voter bias and house effects consist solely of their respective bias terms, namely $\alpha_{2r[i]}\gamma_{g[i]}$ and $\kappa_{h}$. They are also averaged at the election-level, as is average standard deviation, the average election-level value of $\sigma_{i}$ from model~\eqref{poll-model-2}. Details of these calculations are in Appendix~\ref{average-bias}. The estimates of bias in a given column of Table~\ref{tab:model-2-avgs} will not sum to the total (average absolute bias) since they can take different signs. 

The state level aggregation of undecided voters in \eqref{und-model}, used in model \eqref{poll-model-2}, predicts that 3.0\% to 3.8\% of voters were undecided on election day between 2004 and 2012, whilst predicting 5.5\% in 2016. These results are consistent with Figure~\ref{fig:und-roll-ts} which suggested much higher levels of undecideds in 2016 than previous years.

The average absolute (election day) bias in Table~\ref{tab:model-2-avgs} is approximately equal to those in Table~\ref{tab:model-1-avgs}. This indicates that both the original model, and the extended model are able to account for approximately the same amount of bias in the US presidential election polls. However, the extended model is able to disaggregate bias into two additional sources. 

  The average absolute undecided voter bias was estimated to be 2.1\% in 2016. This is more than twice the value in previous years. We can also calculate the average election day bias without undecided voters in 2016, which was only 1.1\% (0.14). This is much closer to the total election day bias estimated in previous years (0.8-1.3\%), and highlights the strong influence undecided voters had on the 2016 presidential election.
  
\begin{figure}
    \centering	\makebox{	\includegraphics[width=0.8\textwidth]{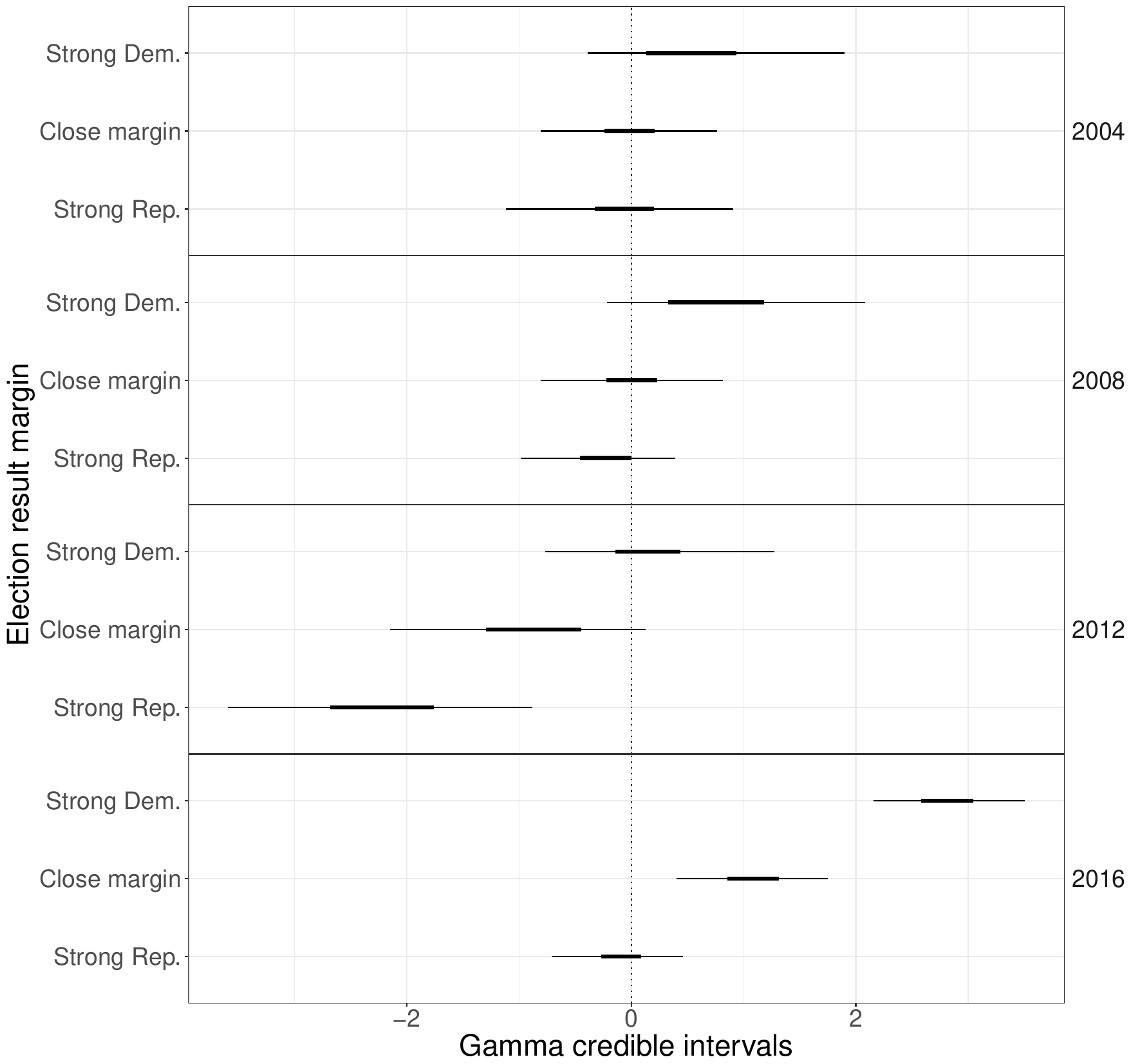}
	}
    \caption{\label{fig:gamma-ci}Credible intervals of 95\% (outer line) and 50\% (inner line) for the effect of undecided voters on polling bias in the model ($\gamma_{g}$) on logit scale. A positive value indicates a bias away from proportional allocation of undecided voters in favour of the Republican candidate. A negative value is favourable for Democratic candidates.}
\end{figure}

\begin{figure}
    \centering	\makebox{	\includegraphics[width=0.8\textwidth]{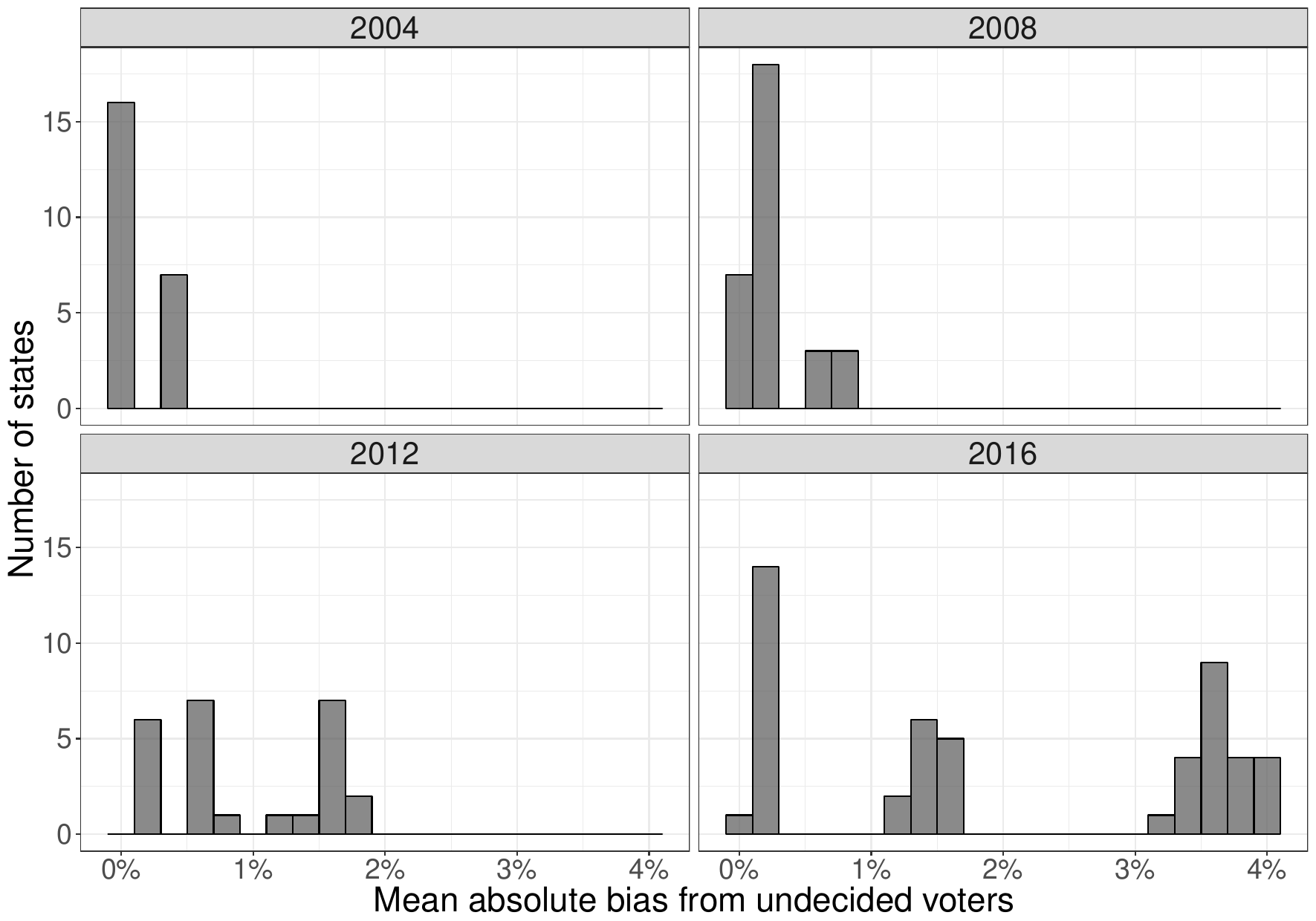}
	}
    \caption{\label{fig:und-abs-bias}Histograms of the average absolute bias from undecided voters for each state-level election, separated by year. The bias from undecided voters is the quantity $\alpha_{2r}\gamma_{g}$ in the model. A positive value indicates a bias away from proportional allocation of undecided voters in favour of either candidate.}
\end{figure}

   The bias attributable to undecided voters in 2004 and 2008 is very small, only 0.3-0.4\%. Since the posterior estimates for the effect of undecideds on bias ($\gamma_{g}$) are mostly centred close to zero (see Figure~\ref{fig:gamma-ci}), and the estimated level of undecided voters was relatively low (Table~\ref{tab:model-2-avgs}), the role of undecided voters seems minimal in these years. The aggregate effect of these factors can be seen in Figure~\ref{fig:und-abs-bias}. In the 2012 election a change occurs when undecided voter bias moves from 0.4\% (2008) to 1.0\%. The potential effect of undecided voters in this year may have been mitigated by the relatively low levels polled (only 3.0\% on average).

\begin{table}
	\caption{\label{tab:model-2-avgs}Average election-level absolute bias and average election-level standard deviation across state-elections in given year(s) from model \eqref{poll-model-2} with assumption of proportional allocation of undecided voters}
  	\centering
	\fbox{%
	\begin{tabular}{l|*{4}{c}|c}
		 & & & & & \em Overall \\ 
		 & 2004 & 2008 & 2012 & 2016 & 2004--2016  \\ \hline
		\multirow{2}{*}{Average absolute bias} & 0.8\% & 1.0\% & 1.3\% & 2.6\% & 1.7\% \\
							  				   & (0.11) & (0.10) & (0.10) & (0.10) & (0.06) \\[0.15cm]  
		\multirow{2}{*}{Average absolute election day bias} & 0.8\% & 0.9\% & 1.3\% & 2.4\% & 1.6\% \\
							  				  			    & (0.12) & (0.11) & (0.14) & (0.12) & (0.07) \\[0.15cm]
		\multirow{2}{*}{Average absolute undecided voter bias} & 0.3\% & 0.4\% & 1.0\% & 2.1\% & 1.1\% \\
							 								   & (0.17) & (0.17) & (0.29) & (0.25) & (0.11) \\[0.15cm]
		\multirow{2}{*}{Average absolute house effects} & 0.6\% & 0.4\% & 0.2\% & 0.2\% & 0.3\% \\
							 							& (0.15) & (0.12) & (0.08) & (0.09) & (0.09) \\[0.15cm]  

		\multirow{2}{*}{Average standard deviation} & 2.2\% & 2.2\% & 2.1\% & 2.4\% & 2.2\% \\
													& (0.04) & (0.04) & (0.04) & (0.05) & (0.03) \\[0.15cm] 
		\multirow{2}{*}{Average election day undecided} & 3.3\% & 3.8\% & 3.0\% & 5.5\% & 4.2\% \\
		  												& (0.24) & (0.21) & (0.21) & (0.28) & (0.14) \\
	\end{tabular}}
\end{table}

To further elicit the role of undecided voters in 2016, Figure~\ref{fig:gamma-ci} contains the 95\% and 50\% credible intervals of the effect size of undecided voters on the polling bias, $\gamma_{g}$, on the logit scale. This parameter vector estimates the biasing effect attributable to undecided voters for twelve groups discussed in Section~\ref{methods:model} and shown in Figure~\ref{fig:und-err-grid}. In terms of the election outcome, we should pay most attention to the close margin category due to the winner-takes-all effect of the Electoral College system. 

The credible intervals of~$\gamma_{g}$ in election years 2004 and 2008 show similar results. Both the strong Republican and close margin categories show no sign of bias from undecided voters. Evidence for polling bias in these years and groups could be mixed (changing from state to state) or inconclusive. There is weak evidence from the strong Democrat category that some bias was induced by the undecided voters (away from proportional allocation) toward the Republican candidate. In 2012, the reverse occurs -- the strong Republican group has bias toward the Democratic candidate. There is some evidence that a biasing effect was present for the close margin group in 2012, but it may also be low or negligible. Figure~\ref{fig:und-err-grid} shows the mean level of undecided voters was relatively low in 2012's close margin category, so overall undecided voters were unlikely have had a substantive effect in 2012. In general, there seems to be a skew toward the Democratic candidate in 2012, whilst the reverse pattern emerges even more strongly in 2016. 

The 95\% credible intervals for 2016's close margin group range from 0.40 to 1.75 with a mean of 1.08. This is the only year (in our set) where the close margin states so clearly induced bias into election polls. Considering estimated undecided numbers averaged 5.5\% on election day in 2016, this was an important source of bias in 2016 and reveals one reason why Trump underperformed in the polls. The strong Democratic group had an even higher bias in 2016 toward Trump, with mean 2.81. This is an interesting result in its own right, (as is 2012's strong Republican category), but had little or no effect on the perception of poll failure since the binary prediction of these races are still likely to be correct, and hence unlikely to change the Electoral College predictions.

Figure~\ref{fig:und-abs-bias} shows the distribution of state level average absolute bias from undecided voters on election day. This figure highlights the consequence of high numbers of undecided voters combined with their large biasing effect in 2016 (and to much less of an extent in 2012). The undecideds' contribution to polling bias in 2016 is made up of three distinct groups, roughly translating to the aforementioned categories.  The effect in 2016 was up to approximately 4\%, whilst the overall effect of undecideds has been almost negligible in 2004 and 2008. The low impact of undecided voters prior to 2016 is likely due to the relatively low levels observed, combined with the possibility that undecided voters did not bias polls cohesively in previous years. 

A house effect ($\kappa_{h}$), attributable to the bias induced by polling agencies methods and practices, is estimated for 39 such firms in the model. Overall, we find a greater number of polling agencies have high bias favouring Republican candidates than high bias towards Democratic candidates, and an overall skew favouring Republicans. However, more polls are needed to accurately assess the Democratic-leaning pollsters. The estimates of this component of bias are presented and discussed further in Appendix~\ref{house-effects}.

 Figure~\ref{fig:state-eld-bias} compares the 2016 election day bias of the polls in each state ($\alpha_{1r} - \alpha_{2r}\gamma_{g}$ in the model) for the two types of static undecided voter allocations (see Appendix~\ref{split-50} for results of even allocations). The grey bars indicate the 95\% credible intervals of election day bias for proportional allocation, whilst the black bars are for even allocation. The state election day biases show that neither allocations performed uniformly better than the other. Proportional allocation resulted in polls with a relatively less biased performance in Utah, Idaho, Iowa, and California for example. Whereas in Vermont, Maine, and Alaska the even allocation was performed marginally better. The relatively larger credible intervals for the even allocation can be attributed to the reduction in sample size for this model which is explained in Appendix~\ref{split-50}. 
 
 Using an even allocation of undecided voters caused no substantive change to the results above, which assumed proportional allocation. The quantitative results in Appendix~\ref{split-50} and Figure~\ref{fig:state-eld-bias} show that both allocation methods were insufficient to incorporate the necessary uncertainty in undecided voter allocation.   In summary, a rigid and non-probabilistic rule for allocating undecided voters was not appropriate for the 2016 US presidential election. 

\begin{figure}
    \centering	\makebox{	\includegraphics[width=0.95\textwidth]{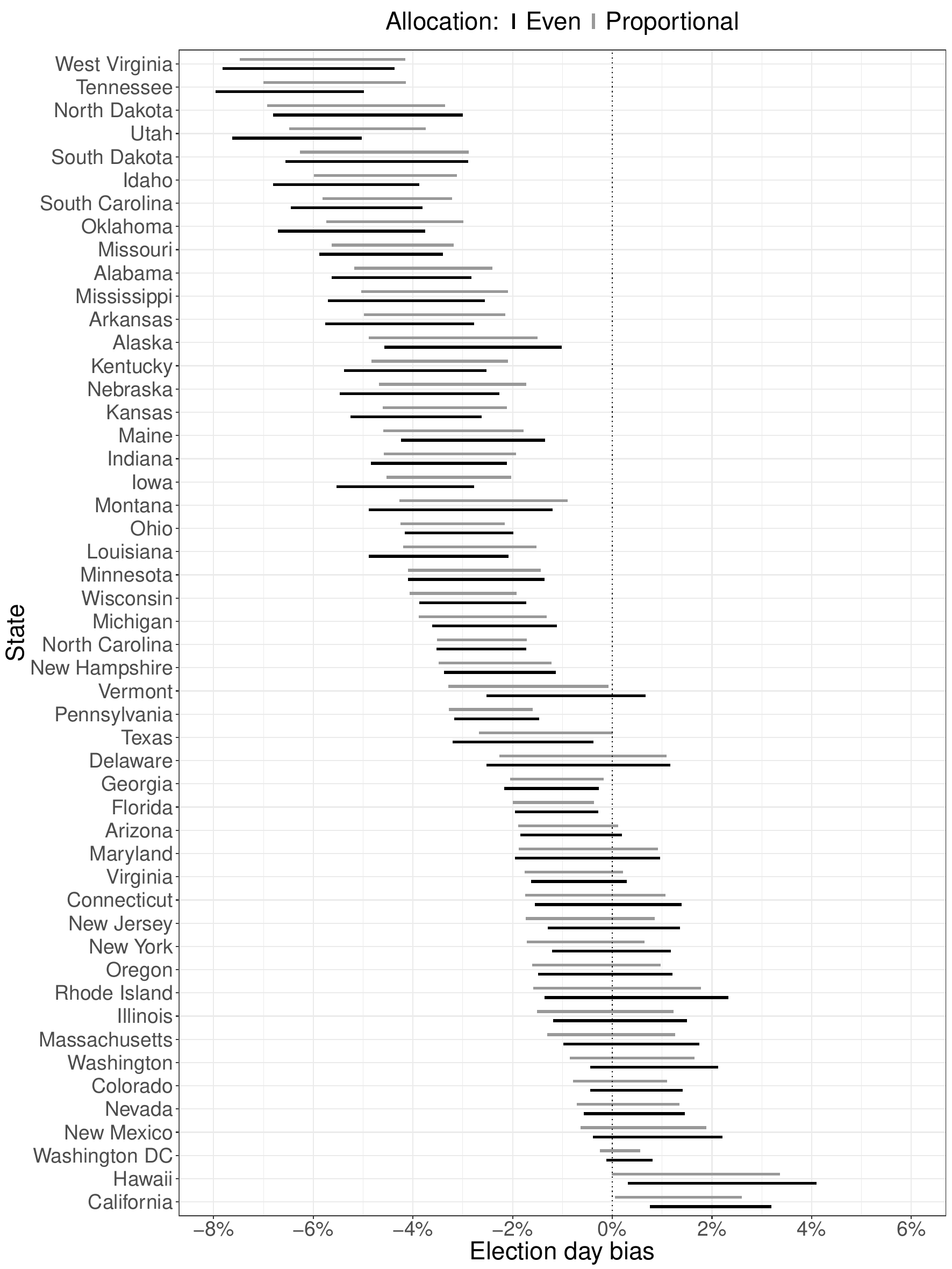}
	}
    \caption{\label{fig:state-eld-bias} 95\% Credible intervals for state election day bias (from true election results) in the 2016 presidential election for proportional (grey line) and even (black line) undecided voter allocation models, after accounting for house effects. A positive value indicates a bias of the polls in favour of the Republican candidate, hence an underperformance in the actual election result.}
\end{figure}  

\section{Discussion}
While there have been methodological advances in election polling and modelling over time, measuring public opinion is still difficult.  Polls have `failed' to accurately predict winning candidates (at least according to the media) in several recent elections, such as the 2015 British election, the Scottish independence referendum, the Brexit referendum, the 2014 US House elections and the 2016 US presidential election (the focus of our paper). Our analyses revealed that one important source of bias in the polls may be undecided voters. We showed that undecided voters biased polls in the 2016 US presidential election by up to 4\% in some states (all things being equal). This bias is particularly problematic given the increasing number of undecided voters observed in our analysis. We found that in 2016, 5.5\% of voters were undecided on election day-- up from 3-4\% in previous years. Others have also found that the percentage of voters who are undecided in the final week of an election campaign is high (up to 30\% in some countries) and may be increasing \citep{Irwin2008,Gelman1993,Orriols2014,Sturgis2016}, although \citet{Sturgis2016} note that the overall effect of late-deciding voters was modest, at most 1\%, in the 2015 British general election.  Given the rising prevalence of indecisive, late-revealing or late-deciding voters, and the bias they may introduce to polling, more attention needs to be given to this group when predicting election outcomes. 

It is well known within the survey research community that, polls suffer from both sampling (due to the fact that information has been collected from a sample rather than everyone in the population) and non-sampling (due to the fact that there is underlying differences in voting behaviour and voting outcomes) errors. Statistical and operational adjustments compensate for sampling errors, but in reality it is the non-sampling errors that play a significant role in the discrepancies between the poll results and election outcomes. The total survey error approach provides a methodology for capturing both types of errors. We have used this approach to understand, interpret and report the various sources of error that in exist in election polling. We have focused specifically on undecided voters, and provide evidence that found that there was substantial differences in the degree of undecidedness in pre-election polling in the 2016 US election. 

Since the majority of polling agencies had no specific methodology to include this in their predictive models, the reported results over-estimated the lead of the Democratic candidate, Clinton, against the Republican candidate, Trump. Our results show that voters who were undecided at the time of being surveyed tended to behave differently to those who decided which party or candidate to vote for earlier. Although it is well recognised that undecided respondents contribute to polling error, there is no consensus about the inferences that can be drawn from their data \citep{Henderson2016,Fenwick1982,Hillygus2011}. Our research has demonstrated that a failure to adequately include them leads to inaccuracies in polling predictions. Specifically, static proportional and even allocations of undecided voters led to bias in polling of the 2016 presidential election. Because of this, we argue that undecided voter counts should always be reported, and probabilistic allocation of undecided voters needs to be implemented in future predictive models. This will allow the uncertainty attributable to undecided voters to propagate through these election models, and improve predictions.

The American Association for Public Opinion Research's (AAPOR) report on US presidential election polling in 2016 \citep{aapor2016} concluded that, in general, polls did not fail relative to historical standards but did under-estimate the support for a Trump presidency in some key states. The authors cited late-deciding voters as one of several explanations for Trump outperforming Clinton on election day. Our study complements AAPOR's report by demonstrating a sizeable number of undecided voters on election day relative to previous elections and by finding a significant effect of undecided voters on the bias in election polls in 2016 using a total survey error model. The AAPOR report relied on exit poll data, whereas we use polling data prior to the election, meaning it may be possible to detect and mitigate these effects in future predictions. We also show that the role of undecided voters in 2016 was different to the 2004, 2008, and 2012 elections. In these previous elections, undecided voters did not significantly affect the races which were close, nor did they significantly contribute to overall polling bias as the number of undecideds remained low. Some bias due to undecided voters was visible in elections prior to 2016, however, the affected states  were not tight races and therefore did not influence the binary outcome of the election. 

Despite the novelty of our findings in ascertaining the role of undecided voters and the adoption of the total survey error framework, our analyses has some noteworthy limitations. Our models remain associational and only provides evidence to support the hypothesis that there is a relationship between the increase in undecidedness and polling accuracy. Properly understanding the underlying effects, and causal mechanisms surrounding how indecision directly influences election outcomes will require future interdisciplinary research. Figure~\ref{fig:gamma-ci} shows that there have been other election years where undecided voters have biased polls away from proportional allocation for certain groups of states. Future research could elicit similarities between these groups which may be useful for predicting the biasing effect of undecided voters. 

In our analysis we would have liked to model more sources of error explicitly, particularly other sources of error highlighted in \citet{aapor2016}. For example, accounting for over-representative sampling of college graduates in some polls would be helpful. However, we were constrained by lack of survey methodology disclosure by polling agencies that limits the number of attributes recorded for polls in our dataset. 

Additionally, though our modelling allows us to quantify the errors that are left unmeasured in standard election level estimates of accuracy, we have not translated this to a model of how respondents will vote in future elections. For those wishing to predict US presidential elections, our analysis demonstrates that if undecided voter levels are high they must be included in the modelling process and that deterministic allocation is not appropriate. For example, an averaging method could model undecideds levels by state, such as in Equation~\eqref{und-model}, and allocations to candidates can be simulated (by Bayesian methods or bootstrapping say). Probabilistic allocation should at least lead to better estimates of uncertainty, and thereby model calibration. The posterior estimates of undecided allocation bias presented in this paper may also be used as a starting place for predictive modellers to incorporate this bias.

Undecided voters played a pivotal role in the 2016 US presidential election, contributing significantly to the bias observed in the polls. We have shown that static allocation methods (proportional and even) were inadequate in the 2016 election. Probabilistic allocation should be considered in future elections, but further investigation and validation of specific methods is needed especially considering the limited supplementary information provided by commercial polls.

\section*{Acknowledgement}
The authors would like to thank the anonymous reviewers for their valuable comments which improved the paper.

\bibliography{und_voters_jrss} 
\clearpage

\appendix
\section{Prior distributions for undecided voter analysis} \label{priors}
The priors used for the polling model, in Table~\ref{tab:priors}, are identical to those from \citet{Shirani2018} expect for the undecided voter allocation, $\gamma_{g}$, and bias from house effects, $\kappa_{h}$, which are new to the model. The notation $\mathcal{N}_{+}(\mu,\sigma^{2})$ denotes half-normal distribution, while $\mathcal{L}(\mu,\sigma)$ represents the Laplace or double-exponential distribution.

The (non-hierarchal) prior probability distribution for $\gamma_{g}$ is double-exponential as we wish to shrink the estimates towards zero where there is insufficient evidence to suggest a true effect exists. We base this prior on Figure~\ref{fig:und-err-grid} where we observe only some categories appearing to demonstrate a relationship between error and the level of undecided voters. The undecided voters on election day, $\alpha_{2r}$, is scaled in the polling model by 10, so that its value is approximately on the same scale as $t_{i}$. This is done for computational reasons, but as a result the prior scale assigned to $\gamma_{g}$ (0.05) is comparable to the scales of other priors. We set its value to be a quarter of the original parameters prior scales to induce further shrinkage.

The house effects bias, $\kappa_{h}$, is specified with a hierarchal prior, similarly to the other hierarchal parameters in the model. Some polling agencies have few polls included in the sample and we would like the parameter estimates to borrow strength across the agencies -- just as in the state-level election day bias parameter, $\alpha_{1r}$. However, we assign a exponential distribution to the standard deviation, $\sigma_{\kappa}$ to induce further shrinkage, relative to $\alpha_{1r}$ and $\beta_{1r}$. Implicitly, we are assuming that some pollsters actually have negligible bias due to their practices so we reflect this in our priors. The scale parameters of this prior are also a quarter of the original parameters prior scales.

The priors for the undecided model are also hierarchal where appropriate. We assign variance levels so that only weak information over the observed undecided voter proportion range (between approximately 0 and 0.2\%) is provided. The prior for $\alpha_{2r}$ has mean, $\phi_{y}$, that varies across year. This is included because the mean of the model is uncentered, unlike the polling model which uses $v_{r}$, and we observe different levels of undecided voters across years.   The prior for $\phi_{y}$, the mean level of undecided voters in each year, is weakly-informative but centred on 0.04. This value is the approximate mean observed in Figure~\ref{fig:und-roll-ts}.

\begin{table}
	\caption{\label{tab:priors}Priors used in models for analysis of state polls.}
  	\centering
	\fbox{%
	\begin{tabular}{*{2}{l}*{3}{c}}
		 & & \em Prior & \multicolumn{2}{c}{\em Hyper-prior}  \\ 
		Model & Component &  & Mean & Variance   \\ \hline
		\multirow{4}{*}{Polling} & \multirow{4}{*}{Mean} & $\alpha_{1r} \sim \mathcal{N}(\mu_{1\alpha}, \sigma_{1\alpha}^{2})$ & $\mu_{1\alpha} \sim \mathcal{N}(0,0.2)$ & $\sigma_{1\alpha}\sim \mathcal{N}_{+}(0,0.2)$ \\ 
			&   & $\beta_{1r} \sim \mathcal{N}(\mu_{1\beta}, \sigma_{1\beta}^{2})$ & $\mu_{1\beta} \sim \mathcal{N}(0,0.2)$ & $\sigma_{1\beta}\sim \mathcal{N}_{+}(0,0.2)$\\
			&   & $\gamma_{g} \sim \mathcal{L}(0,0.05)$ &   &   \\
			&   & $\kappa_{h} \sim \mathcal{N}(\mu_{\kappa},\sigma_{\kappa}^{2})$ & $\mu_{\kappa} \sim \mathcal{N}(0,0.05)$  & $\sigma_{\kappa} \sim \exp(1/0.05)$ \\
			& Variance & $\tau^{2}_{1r} \sim \mathcal{N}_{+}(0,\sigma_{1\tau}^{2})$ &   & $\sigma_{1\tau} \sim \mathcal{N}_{+}(0,0.05)$  \\
			& & & &  \\
		\multirow{3}{*}{Undecided voters} & \multirow{2}{*}{Mean} & $\alpha_{2r} \sim \mathcal{N}(\phi_{y[r]}, \sigma_{2\alpha}^{2})$ & $\phi_{y} \sim \mathcal{N}(0.04,0.01)$ & $\sigma_{2\alpha}\sim \mathcal{N}_{+}(0,0.02)$ \\
			&  & $\beta_{2r} \sim \mathcal{N}(\mu_{2\beta}, \sigma_{2\beta}^{2})$ & $\mu_{2\beta} \sim \mathcal{N}(0,0.02)$ & $\sigma_{2\beta}\sim \mathcal{N}_{+}(0,0.02)$\\
		 &  Variance & $\tau_{2r}^{2} \sim \mathcal{N}_{+}(0,\sigma_{2\tau}^{2})$ &   & $\sigma_{2\tau} \sim \mathcal{N}_{+}(0,0.01)$  \\
	\end{tabular}}
\end{table}
	 
\section{Mean election-level bias and variance calculations}
	\label{average-bias}
The definitions of mean bias and variance quantities presented in Table~\ref{tab:model-1-avgs}, Table~\ref{tab:model-2-avgs}, and Figure~\ref{fig:und-abs-bias} are taken from \citet{Shirani2018} where appropriate. We restate them here, and add additional definitions for bias from undecided voters and house effects. For a given election race, the average bias, $b_{r}$, is defined as
\begin{equation}\label{all-bias-election-level}
 b_{r} = \frac{1}{|S_{r}|} \sum_{i \in S_{r}} \left(p_{i} - v_{r} \right)
\end{equation}
where $S_{r}$ is the set of polls from state-level election $r$, $|S_{r}|$ denotes the size of the set $S_{r}$, and $p_{i}$ is taken from either model~\eqref{poll-model-1} or~\eqref{poll-model-2} depending on which is in use. Calculating the mean of $\vert b_{r} \vert $ over a given set of elections determines the average election-level absolute bias over that set. The set of races, for example, could be one election year (e.g. all state-level races in 2004), or all of the elections in the dataset.

The bias on election day can be calculated by setting $t=0$ in $p_{i}$ and calculating the same quantity above. More specifically
\begin{equation}\label{eld-bias-election-level}
 b^{e}_{r} = \frac{1}{|S_{r}|} \sum_{i \in S_{r}} \left(p^{e}_{i} - v_{r} \right)
\end{equation}
where $\text{logit}(p_{i}^{e}) = \text{logit}(v_{r[i]}) + \alpha_{1r[i]} - \alpha_{2r[i]}\gamma_{g[i]} + \kappa_{h[i]}$.
Similarly, bias attributable to certain components or excluding components can be calculated by setting the unwanted parameters in $p_{i}$ to zero. The full list of bias measures are in defined at the poll-level Table~\ref{tab:bvquants}. As in Equations~\eqref{all-bias-election-level} and~\eqref{eld-bias-election-level}, the election-level bias for each race $r$ can be calculate by taking the mean over the polls of that election.  

Finally, the average absolute election-level bias, $\mu_{b,S}$, of type $b$, over subset of elections $S$, can be calculated by
\begin{equation}
	\mu_{b,S} = \frac{1}{|S|} \sum_{s \in S} \vert b_{s} \vert.
\end{equation} 

\begin{table}
	\caption{\label{tab:bvquants} Descriptions and definitions of poll-level bias quantities used to calculate election-level bias (and averages) on percent scale.}
  	\centering
	\fbox{%
	\begin{tabular}{l|l|l}
	Name & Sym. & Poll-level metric \\ \hline
	Model~\eqref{poll-model-1} & & \\
	\quad All bias & $b_{r}$ & $\text{logit}(p_{i}) = \text{logit}(v_{r[i]}) + \alpha_{1r[i]} + t_{i}\beta_{1r[i]}$ \\
	\quad Election day bias & $b_{r}^{e}$ & $\text{logit}(p_{i}^{e}) = \text{logit}(v_{r[i]}) + \alpha_{1r[i]}$ \\
	Model~\eqref{poll-model-2} & & \\
	\quad All bias & $b_{r}$ & $\text{logit}(p_{i}) = \text{logit}(v_{r[i]}) + \alpha_{1r[i]} + t_{i}\beta_{1r[i]}- \alpha_{2r[i]}\gamma_{g[i]} + \kappa_{h[i]}$ \\

	\quad Election day bias & $b_{r}^{e}$ & $\text{logit}(p_{i}^{e}) = \text{logit}(v_{r[i]}) + \alpha_{1r[i]} - \alpha_{2r[i]}\gamma_{g[i]} + \kappa_{h[i]}$ \\
	\quad Undecided voter bias & $b_{r}^{u}$ & $\text{logit}(p^{u}_{i}) = \text{logit}(v_{r[i]}) - \alpha_{2r[i]}\gamma_{g[i]}$ \\
	\quad House effects & $b_{r}^{h}$ & $\text{logit}(p^{h}_{i}) = \text{logit}(v_{r[i]}) + \kappa_{h[i]}$

	\end{tabular}}
\end{table}
 
The average election-level polling standard deviation is defined similarly by
\begin{equation}
	\sigma_{r} = \frac{1}{|S_{r}|} \sum_{i \in S_{r}} \sqrt{\frac{p_{i}(1-p_{i})}{n_{i}} + \tau^{2}_{1r}}
\end{equation}
and the average over a set of elections can be calculated based on this quantity also.

\section{Bias from house effects} \label{house-effects}
The house effect for a given polling organisation is the bias attributable to their specific polling methods and practices. Unfortunately, we are only able to capture an aggregate of house polling bias, because polling agencies only release a small amount of information on the assumptions behind their polls. Of the total number of polls, 85\% have an associated house effect estimated. Polls without an associated house effect are due to insufficient poll numbers from the pollster. 

\begin{figure}
    \centering	\makebox{	\includegraphics[width=0.95\textwidth]{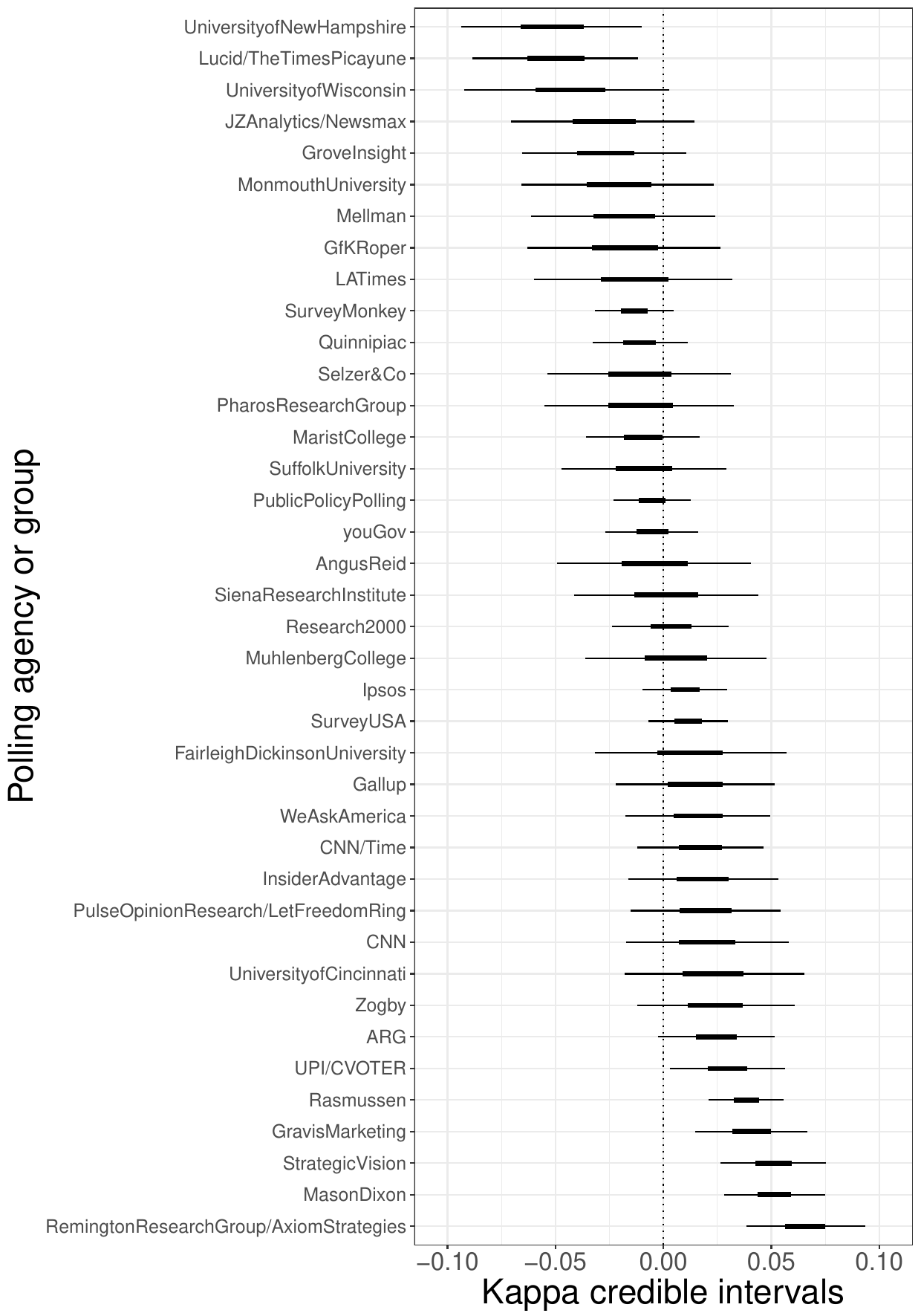}
	}
    \caption{\label{fig:kappa-ci-ext} Credible intervals of 95\% (outer line) and 50\% (inner line) for the house effects bias from polling organisations in the model ($\kappa_{h}$), on the logit scale. A positive value indicates a bias in favour of the Republican candidate, hence underperformance in the election result.}
\end{figure}

Figure~\ref{fig:kappa-ci-ext} shows some interesting trends in partisanship and polling. Overall, there is a small skew toward pollsters being biased in favour of Republican candidates, and far more of these Republican-biased agencies are decisively biased. Rassmussen, Gravis Marketing, Mason-Dixon, Strategic Vision, and Remington Research Group/Axiom Strategies have estimated average bias between 0.95\% and 1.64\% on average (Table~\ref{tab:houseeffects}). On the other hand, the University of New Hampshire and Lucid/The Times Picayune appear to have substantial bias in favour of the Democratic candidate -- over 1.2\% is shown in Table~\ref{tab:houseeffects}. The wider credible intervals, relative to their Republic-bias counterparts, may be due to a lower number of polls included in the data set (18 and 13 respectively, versus between 24 and 198 polls per group). 

\begin{table}
	\caption{\label{tab:houseeffects} Average house effects across elections. Only those polling agencies with absolute mean posterior greater than 0.5\% are shown.}
  	\centering
	\fbox{%
\begin{tabular}{lrr}
  \hline
  & \multicolumn{2}{c}{\em Posterior} \\
Polling agency or group & mean & s.d. \\ 
  \hline
ARG & 0.61 & 0.35 \\ 
CNN & 0.50 & 0.48 \\ 
Gravis Marketing & 1.01 & 0.33 \\ 
Grove Insight & -0.67 & 0.49 \\ 
JZ Analytics / Newsmax & -0.69 & 0.54 \\ 
Lucid / The Times Picayune & -1.23 & 0.49 \\ 
Mason Dixon & 1.27 & 0.29 \\ 
Monmouth University & -0.51 & 0.55 \\ 
Rasmussen & 0.95 & 0.22 \\ 
Remington Research Group / AxiomStrategies & 1.64 & 0.35 \\ 
Strategic Vision & 1.27 & 0.31 \\ 
University of Cincinnati & 0.58 & 0.53 \\ 
University of New Hampshire & -1.28 & 0.53 \\ 
University of Wisconsin & -1.06 & 0.59 \\ 
UPI/CVOTER & 0.69 & 0.32 \\ 
Zogby & 0.60 & 0.46 \\ 
   \hline
\end{tabular}}
\end{table}

Table~\ref{tab:houseeffects} shows selected estimates of the overall bias, $b_{h}$, attributable to pollster $h$ on the percentage scale. These values are calculated by
\begin{equation}
		 b_{h} = \frac{1}{|S_{h}|} \sum_{i \in S_{h}} \left(p^{h}_{i} - v_{r[i]} \right)
\end{equation}
where $S_{h}$ is the set of polls taken by pollster $h$, $p^{h}_{i}$ is the poll-level house effect bias (see Table~\ref{tab:bvquants}), and $v_{r[i]}$ is the election result in race $r$ associated with poll $i$.

The house effect bias of a particular agency may be unintentional and unassociated with partisanship. For example, the relative bias toward Democratic candidates by Survey Monkey could be a result of their online sampling frame which is likely skewed toward a younger demographic. Most, if not all, organisation use methods to mitigate this bias but it is not alway possible when limited demographics are sampled.  

The pollsters with arguably the best performance are youGov and Public Policy Polling. However, both have a relatively high number of polls included in the dataset, 63 and 106 each. The average number of polls per pollster is 41, and range of polls is 8 to 184. 

\section{Modelling with an even allocation of undecided voters} \label{split-50}
In order to embed an even allocation of undecided voters into model~\eqref{poll-model-2}, we make the following changes.
\begin{enumerate}
	\item We remove the polls that do not report an undecided voter number as, unlike the proportional case, we cannot infer the even allocation from the data. Following their removal, 1,725 polls remain in the dataset. 
	\item The poll represented in~\eqref{prop-poll-alloc} changes so that the poll values are scaled to remove third party candidates then Republicans are allocated 50\% of the undecided voters. Formally, this can be represented most succinctly by 
	\begin{equation*} 
		y_{i} = \frac{R_{i} + 0.5 \times U_{i}}{R_{i} + D_{i} + U_{i}}.
	\end{equation*}
	\item The undecided voter allocation with uncertainty, $\lambda$ in \eqref{uncertain-alloc}, changes to $\lambda = 0.5 + \theta_{i}$.
\end{enumerate}
Given the updated definitions of $y_{i}$ and $\lambda$, using \eqref{scale-3p} we note that the identity in \eqref{prop-alloc-identity} still holds and modelling can proceed as in the case of a proportional allocation of undecided voters.

The analysis with an even allocation of undecided voters was very similar to the original results in Section~\ref{sec:results}. In general, most measures of bias across the years slightly increased, and the average standard deviation in the model increased. This is indicative of a bias-variance trade-off in the modelling process. Overall, summary measures for elections years 2004--2016 had an increase in the bias of 0.1\% (average absolute bias and election day bias), and a decrease in the standard deviation of 0.1\%. Focussing on 2016, average election day bias increased by 0.2\%, whilst the average standard deviation decreased by 0.2\%. 

 \begin{table}
	\caption{\label{tab:model-2-avgs-50}Average election-level absolute bias and average election-level standard deviation across state-elections in given year(s) from model \eqref{poll-model-2} with assumption of even allocation of undecided voters.}
  	\centering
	\fbox{%
	\begin{tabular}{l|*{4}{c}|c}
		 & & & & & \em Overall \\ 
		 & 2004 & 2008 & 2012 & 2016 & 2004--2016  \\ \hline
		\multirow{2}{*}{Average absolute bias} & 0.8\% & 1.1\% & 1.4\% & 2.8\% & 1.8\% \\
							  				   & (0.11) & (0.10) & (0.11) & (0.10) & (0.06) \\[0.15cm]  
		\multirow{2}{*}{Average absolute election day bias} & 0.8\% & 1.0\% & 1.4\% & 2.6\% & 1.7\% \\
							  				  			    & (0.12) & (0.12) & (0.14) & (0.12) & (0.07) \\[0.15cm]
		\multirow{2}{*}{Average absolute undecided voter bias} & 0.4\% & 0.5\% & 1.0\% & 2.3\% & 1.3\% \\
							 								   & (0.19) & (0.21) & (0.30) & (0.24) & (0.11) \\[0.15cm]
		\multirow{2}{*}{Average absolute house effects} & 0.6\% & 0.4\% & 0.3\% & 0.2\% & 0.4\% \\
							 							& (0.17) & (0.14) & (0.12) & (0.06) & (0.09) \\[0.15cm]  

		\multirow{2}{*}{Average standard deviation} & 2.1\% & 2.1\% & 2.0\% & 2.2\% & 2.1\% \\
													& (0.04) & (0.03) & (0.03) & (0.04) & (0.03) \\[0.15cm] 
		\multirow{2}{*}{Average election day undecided} & 3.3\% & 3.8\% & 3.0\% & 5.5\% & 4.2\% \\
		  												& (0.24) & (0.21) & (0.21) & (0.28) & (0.14) \\
	\end{tabular}}
\end{table}	

\begin{figure}
    \centering	\makebox{	\includegraphics[width=0.8\textwidth]{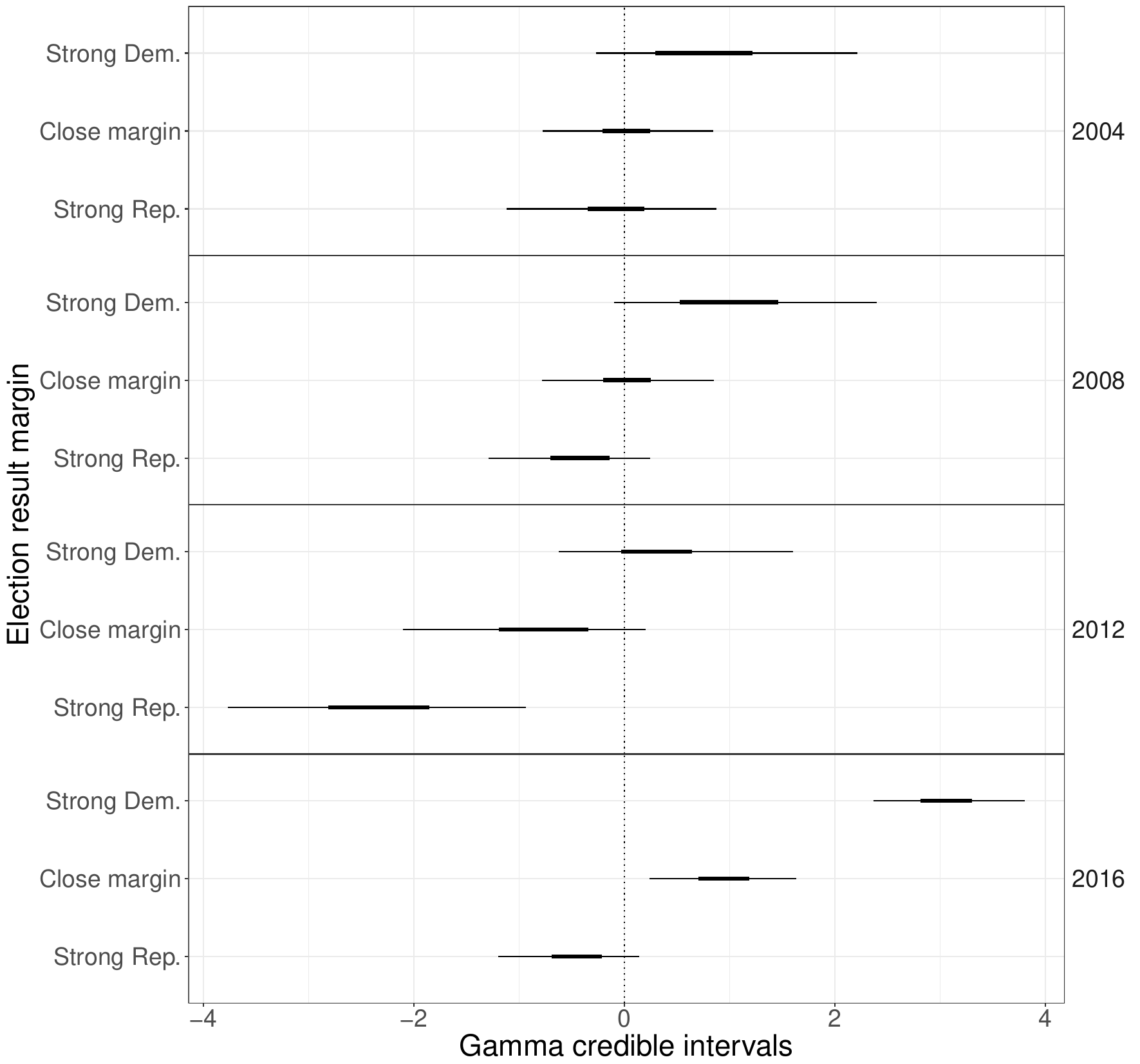}
	}
    \caption{\label{fig:gamma-ci-50}Credible intervals of 95\% (outer line) and 50\% (inner line) for the effect of undecided voters on polling bias in the model ($\gamma_{g}$), on logit scale. A positive value indicates a bias away from an even allocation of undecided voters in favour of the Republican candidate.}
\end{figure}

The effect of undecided voters on the polling bias in the model ($\gamma_{g}$) is replotted in Figure~\ref{fig:gamma-ci-50} with credible intervals. This plot is largely unchanged and still shows that undecided voters in the 2016 election year had a considerably different effect on bias compared to previous years.

\begin{figure}
    \centering	\makebox{	\includegraphics[width=0.8\textwidth]{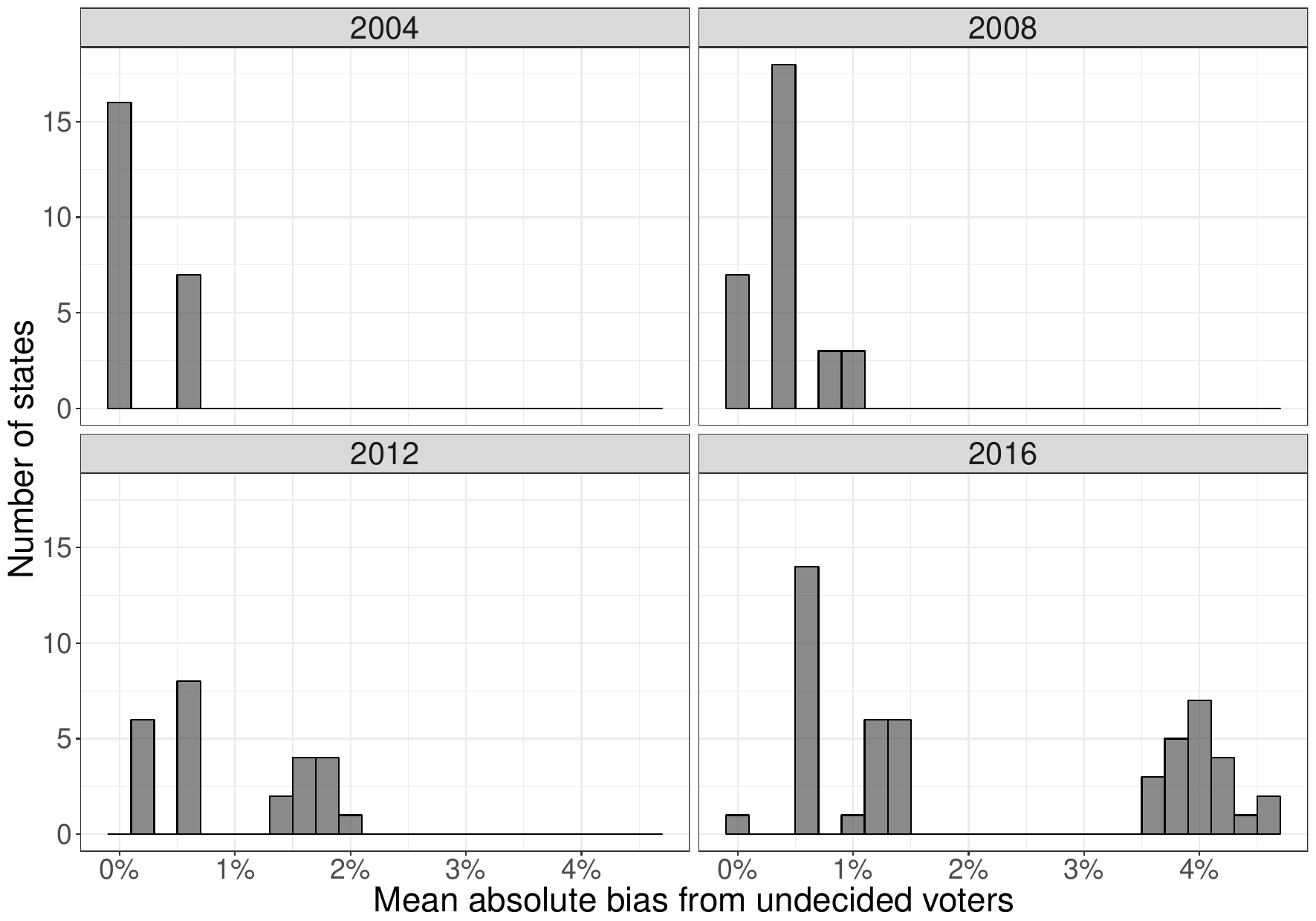}
	}
    \caption{\label{fig:und-abs-bias-50}Histograms of the average absolute bias from undecided voters for each state-level election, separated by year. The bias from undecided voters is the quantity $\alpha_{2r}\gamma_{g}$ in the model. A positive value indicates a bias away from an even allocation of undecided voters in favour of either candidate.}
\end{figure}

The histogram of the absolute bias from undecided voters is replotted for each state-level election in Figure~\ref{fig:und-abs-bias-50} under the assumption of even allocation of undecided voters. It shows a similar story as before. With careful comparison to Figure~\ref{fig:und-abs-bias}, one can see that bias has increased for many of the states. 
	
\end{document}